\documentstyle[prb,twocolumn,aps,epsf]{revtex}
\begin{document}

\twocolumn[
\hsize\textwidth\columnwidth\hsize\csname@twocolumnfalse\endcsname

\title{Correlation induced phonon softening 
in low density coupled bilayer systems}
\author{E. H.\ Hwang and S. Das Sarma}
\address{Department of Physics, University of Maryland, College Park,
Maryland  20742-4111 } 
\date{\today}
\maketitle

\begin{abstract}

We predict a possible phonon softening instability in strongly 
correlated coupled semiconductor bilayer systems. 
By studying the plasmon-phonon coupling in coupled bilayer structures,
we find that the renormalized acoustic phonon frequency may be 
softened at a finite wave vector 
due to many-body local field corrections,
particularly in low density systems where correlation effects are 
strong. We discuss experimental possibilities to search for this 
predicted phonon softening phenomenon.


\end{abstract}
\vspace{0.2in}
]

\newpage

Much interest has recently \cite{ei,das,spi} focused on bilayer 
semiconductor structures, where two quantum wells are put in close 
proximity with a high insulating barrier in-between to suppress 
interlayer tunneling. Such systems are of intrinsic interest because 
competition between intra- and inter-layer Coulomb interaction (i.e., 
correlation effects), particularly at low carrier densities when 
Coulomb interaction dominates (or in high magnetic fields 
\cite{ei,das,spi} when the electronic kinetic energy is quenched), 
may lead to novel and exotic ground states, collective properties, and 
quantum phase transitions. Although much of the recent work has 
concentrated on bilayer quantum Hall systems \cite{ei,das,spi} the 
basic issue of the competition among electron kinetic energy and 
intra-/inter-layer Coulomb correlations is quite generic, and 
interesting novel ground states and exotic collective properties are, 
in principle, possible even in the zero field situation provided the 
$r_s$ parameter defining the dimensionless ratio of Coulomb interaction 
strength to Fermi energy is high enough. In this Letter we predict a 
particularly novel possibility, namely a finite wave vector acoustic 
phonon softening, driven by correlation effects in low density bilayer 
systems. We believe that our predicted phonon softening phenomenon may 
be realizable in the existing (electron or hole) bilayer systems 
provided the carrier density (and temperature) are low enough.

Advances in nanofabrication and growth of semiconductor layered structures
have recently made 
it possible to fabricate very high quality (and very low density)
GaAs-based high-mobility double well (bilayer) two 
dimensional  electron systems (2DES).
The carrier density in these systems could be very low (large $r_s$), 
making them suitable for studying electronic correlation 
in low dimensions.
As the density decreases, unusual 
behavior is expected \cite{neilson,hu,dsh,dassarma,pinczuk}
due to the increasing importance of
electron-electron interaction as the system goes 
from a weak-coupling (small $r_s$) to a strong-coupling (large $r_s$) 
regime. It is possible to have electron systems with $r_s$ as high as 
15 -- 20 and hole systems with $r_s$ as high as 25 -- 30.
One particularly interesting possibility \cite{neilson} is that 
an electronic charge density wave (CDW) instability may occur
in these low density bilayer systems 
before the onset of Wigner crystallization at still lower density.

In bilayer structures  there are two collective charge density 
excitations (plasmon modes) \cite{dassarma,pinczuk}
corresponding to
the in-phase density fluctuation (optic plasmon, OP)
and out-of-phase density fluctuation (acoustic plasmon, AP) between
the layers.
In the absence of any interlayer  tunneling the AP mode has a linear 
dispersion in the 2D wave vector
because the long-range Coulomb interaction is
suppressed by the screening from adjacent layers.
Within random phase approximation (RPA) the acoustic plasmon is
shown to exist for small wave vector (long wavelength) at all densities. 
However, the linear dispersion relation of the acoustic plasmon is 
sensitive to many-electron correlation effects 
even in the long wavelength regime.
It has been shown that electron
correlations may strongly renormalize the AP mode 
by pushing it completely inside the
single-particle electron-hole 
excitation (SPE) spectra, resulting in Landau damping
of the mode at much smaller wave vectors than predicted by 
RPA\cite{neilson,dsh}. Thus, the AP mode may become overdamped due to 
correlation effects.
Strong correlations can thus destroy the 
AP mode without smearing out of 
the boundary of the particle-hole 
continuum by defect or impurity scattering or thermal effects \cite{neilson}.
When this happens, 
there is a strong shift of single particle spectral 
weight toward zero energy at some finite wave vector 
due to the disappearance of the AP mode. 
The appearance of these spectral 
features with zero energy spectral weight at finite wave vector 
indicates that a state with a periodic modulation of 
density costs relatively little energy to excite,
and the system may therefore become unstable against the formation of 
a periodic bilayer CDW 
ground state \cite{neilson} due to this AP plasmon 
overdamping mechanism. 
This CDW is different from the one associated with the 2$k_F$ 
($k_F$ being the Fermi wave vector) Peierls instability
in one dimensional (1D) systems \cite{gruner,senna}, 
where the CDW instability occurs due to 
the electron-acoustic phonon interaction leading to a
strongly divergent 2$k_F$ response function.
The bilayer correlation induced 
CDW instability does not happen at 2$k_F$
or any other Fermi surface related wave vector.

In this Letter we study the electron-acoustic phonon interaction 
in low density bilayer systems.
We investigate the plasmon (both OP and AP) mode coupling 
to the acoustic-phonons of the system  and
the consequent possibility of a 
softening of acoustic phonon modes caused by the 
electron-phonon interaction in 2D bilayer systems. 
The plasmon-phonon 
mode coupling phenomenon, which hybridizes the collective plasmon
modes of the electron gas with the acoustic-phonon modes, gives rise
to coupled plasmon-phonon modes. The mode coupling 
is particularly strong at wave vectors around the 
plasmon-phonon resonance 
condition ($\omega_{pl} \approx \omega_{ph}$). 
In 2D GaAs systems the resonant mode-coupling phenomenon normally occurs
at an extremely low density ($n < 1.2\times 10^8$ $cm^{-2}$) 
in the absence of correlation effects because the acoustic phonon energy 
is typically very low and one has to go to extremely low densities for 
the plasma frequency to be low enough to satisfy the 
$\omega_{pl} \approx \omega_{ph}$ condition.
Our important finding is that in the strongly correlated bilayer systems the
plasmon-phonon mode coupling mechanism is sufficiently enhanced to 
give rise to strong (resonant) mode coupling effect at rather high densities
which are already obtainable in currently
available samples ($n \le 1.0\times 10^{10}$ $cm^{-2}$).
In addition, resonant mode-coupling occurs
at large wave vectors ($q >2 k_F$)
making it potentially experimentally observable.
The mode coupling effect depends strongly on electron density
and layer separation since it originates from Coulomb correlations.
We find that bilayer correlation effects in currently available samples
could be strong enough that mode coupling may lead to a 
renormalized acoustic phonon 
mode which becomes soft at the wave vector around
the resonance. The bilayer charge density
instability associated with the overdamping of AP mode due to
strong electron correlations
gives rise to the drastic modification of 
the acoustic phonon dispersion leading to be softening.

In this calculation we incorporate electron correlation effects 
in the local-field corrections which are
deduced from quantum Monte Carlo numerical 
simulation data for the ground state of the 2DES \cite{tanatar}.
We note that the specific correlation approximations used in 
our theory for the local field correction do not
affect our qualitative conclusion about the existence of phonon 
softening in the low density bilayer systems, although the quantitative 
details (precise density or wave vector at which softening occurs)
depend on the approximation scheme.
We take the bilayer system to be made of two 
identical symmetric quantum wells of width
$a$ separated by a distance $d$.
The Hamiltonian describing the bilayer electron-phonon 
system is given by
$H = H_0 + H_1$, where $H_0$ is the noninteracting electron and 
acoustic phonon kinetic energy, 
\begin{equation}
H_0 = \sum_{k}\sum_{l=1,2}\varepsilon(k) a_{k,l}^{\dagger}a_{k,l}
  + \sum_{Q} \omega_Q b_Q^{\dagger} b_Q,
\end{equation}
and $H_1$ is the electron-electron and the electron-phonon interaction,
\begin{eqnarray}
H_1 &=& \sum_{k,k',q}\sum_{l,l'=1,2}V_{ll'}(q)a_{k+q,l}^{\dagger}
a_{k'-q,l'}^{\dagger}a_{p',l'}a_{p,l} \nonumber \\
& + & \sum_{k,Q}\sum_{l=1,2}M_{l}(Q)
f_l(q_z)a_{k-q,l}^{\dagger}a_{k,l}(b_{Q}^{\dagger} + b_{-Q}),
\label{h1}
\end{eqnarray}
where $a_{k,l}^{\dagger}$, $a_{k,l}$ are the $l$-th layer 
electron creation and destruction operators with 2D wave vector $k$,
$b_{Q}^{\dagger}$ the phonon creation operator with 3D wave vector 
$Q = (q,q_z)$,
$\varepsilon(k) = k^2/2m$ the kinetic energy of the electron
(we take $\hbar =1$ throughout this paper), and
$ \omega_Q$  the bulk phonon frequency. 
$V_{ll'}(q)$ is the Coulomb interaction between $l$-th and $l'$-th layer
whose matrix elements 
are given by $V_{ll'}(q) = v(q) F_{ll'}(q)$, where 
$v_(q)= 2\pi e^2/(\kappa q)$ with $\kappa$ as the
background lattice dielectric constant and $F_{ll'}(q)$ are the form 
factors describing the finite well width effects. 
We consider only the lowest subband (for an infinite square 
well potential) in each layer for simplicity.
For two identical layers we have $V_{11} = V_{22}$ and $V_{12} = V_{21}$.
In Eq. (\ref{h1}), $f_l(q_z) = \int dz \phi_l(z)^* \phi_l(z) e^{iq_z z}$,
where $\phi_l(z)$ is the subband wave function in $l$-th layer.
$M_{l}(Q)$ is the coupling constant between the $l$-th layer electron 
and bulk GaAs phonons. In the jellium model
$M_{l}(Q) = i({2\pi e^2 \omega_Q}/{Q^2})^{1/2}$.
In GaAs/Al$_x$Ga$_{1-x}$As systems, electrons couple to acoustic phonons 
via deformation potential and piezoelectric couplings. 
We consider only the deformation potential coupling, which
dominates in the present context.
The electron-acoustic phonon coupling constant via the 
deformation-potential interaction is given by
$|M_{l}(Q)|^2 = {D^2 Q^2}/(2 \rho \omega_Q)$
where $\rho$ is the mass density of the lattice, $D$  the 
known deformation-potential coupling constant,
and $\omega_Q = c_0 Q$, where $c_{0}$  is the longitudinal 
bulk sound velocity, is the acoustic phonon frequency.

Electrons interact among themselves through the direct Coulomb interaction
and also through the dynamical virtual-acoustic-phonon exchange process. 
The effective interaction of an electron in $l$-th layer with an 
electron in $l'$-th layer is obtained in RPA by summing all the bubble
diagrams
\begin{equation}
W_{ll'}(q,\omega) = \psi_{ll'}(q,\omega) 
+ \sum_{m}\psi_{lm}\Pi(q,\omega) W_{ml'}(q,\omega),
\label{wll}
\end{equation}
where $\psi_{ll'}(q,\omega)=V_{ll'}(q) + U_{ll'}(q,\omega)$, 
and the $U_{ll'}(q,\omega)$ is the phonon-mediated electron-electron 
interaction which is given by summing over the phonon wave vector in the
$z$-direction (the 2D layers are in the $x$-$y$ plane):
\begin{equation}
U_{ll'}(q,\omega)=\sum_{q_z}M_{l}(q,q_z)M_{l'}^*(q,q_z)|f_l(q_z)|^2
D^0(Q,\omega),
\label{ull0}
\end{equation}
where $D^0(Q,\omega) = {2 \omega_Q}/(\omega^2 - \omega_Q^2 +i \gamma)$
is the free-phonon propagator.
We note that except for the form factor $|f(q_z)|^2$ the integrand 
in Eq. (\ref{ull0}) does not 
vary rapidly with $q_z$. We therefore crudely approximate it 
with that factor multiplied by the integrand at $q_z=0$. 
(Explicit calculations show that this approximation is excellent.)
Then we get 
\begin{equation}
U_{ll'}(q,\omega)=\frac{U_0(q)}{s(q,\omega)} \left [
\theta(\omega-\omega_q)J_{ll'}(qs)
-\theta(\omega_q-\omega)I_{ll'}(qs) 
\right ],
\label{ull}
\end{equation}
where $s(q,\omega) = \sqrt{|1-\omega^2/\omega_q^2|}$, and
$U_0(q) = D^2q^2/(2\rho c_0^2)$ for the deformation potential 
and $U_0(q)=v(q)$ for the jellium model, respectively, and
$I_{ll'}(q)  = \int dz dz' |\phi_l|^2 |\phi_{l'}|^2 
e^{-|z-z'|q}$, 
$J_{ll'}(q)  = \int dz dz' |\phi_l|^2 |\phi_{l'}|^2 e^{-i|z-z'|q}$.

The effective electron-electron interaction matrix, Eq (\ref{wll}),
can be diagonalized through a canonical transformation by 
introducing symmetric-antisymmetric (SAS) operators 
($\alpha_{k,\pm} = [a_{k,1} \pm a_{k,2}]/\sqrt{2}$) in the Hamiltonian. 
In the SAS representation the effective interaction then
becomes decoupled by virtue of the symmetric nature of the bilayer
system. The 
diagonalized elements of the effective interaction are given by
\begin{equation}
W_{\pm}(q,\omega)  = \frac{\psi_{\pm}(q,\omega)}{1 +
\psi_{\pm}(q,\omega) \Pi(q,\omega)} 
 =  \frac{V_{\pm}(q)}{\epsilon^{tot}_{\pm}(q,\omega)},
\label{etot}
\end{equation}
where the plus (minus) label corresponds to the in-phase (out-of-phase)
charge density modulation of the two layers. 
$\psi_{\pm}(q,\omega) = V_{\pm}+U_{\pm}$, where
$V_{\pm}(q) = V_{11}(q)[1-G_{11}(q)] \pm V_{12}(q)[1-G_{12}(q)]$ is
the Coulomb interaction matrix elements with the proper intralayer and 
interlayer local field corrections $G_{11,12}(q)$ \cite{neilson,tanatar}; 
and $U_{\pm}(q,\omega) = U_{11}(q,\omega) \pm U_{12}(q,\omega)$ is the 
phonon mediated interaction 
matrix element; and $\Pi(q,\omega)$  the
noninteracting 2D polarizability function. 
$\epsilon^{tot}_{\pm}(q,\omega)$ is the total dielectric function of the 
coupled electron-phonon system and is given by
\begin{equation}
\epsilon^{tot}_{\pm}(q,\omega) = 1 + V_{\pm}(q)\Pi(q,\omega) 
-\frac{U_{\pm}(q,\omega)}
{V_{\pm}(q) + U_{\pm}(q,\omega)}.
\end{equation}
The dispersion relation of the coupled plasmon-acoustic phonon mode 
is given by the zeros of the dielectric function 
$\epsilon^{tot}_{\pm}(q,\omega)$. We can rewrite Eq. (\ref{etot}) in terms
of the screened Coulomb interaction and renormalized phonon interaction
\begin{equation}
W_{\pm}(q,\omega) = \frac{V_{\pm}(q)}{\epsilon_{\pm}(q,\omega)}
+ \frac{M_{\pm}^2(q,\omega)}{\epsilon_{\pm}^2(q,\omega)} D_{\pm}(q,\omega).
\label{wpm}
\end{equation}
The first term is the screened Coulomb interaction due to the collective 
motion of the electrons, and the second term represents the renormalized
electron-phonon interaction with the renormalized vertex 
$M_{\pm}/\epsilon_{\pm}$ and phonon propagator $D_{\pm}$.
$\epsilon_{\pm}(q,\omega)$ is the symmetric (antisymmetric) dielectric 
function of the electron system and is given by
$\epsilon_{\pm}(q,\omega) = 1- V_{\pm}(q) \Pi(q,\omega)$.
The renormalized phonon propagator $D_{\pm}(q,\omega)$ is
\begin{equation}
D_{\pm}(q,\omega) = \frac{2 \omega_q}{\omega^2 - \omega_q^2 +
2\omega_q M_{\pm}^2(q,\omega) \chi_{\pm}
(q,\omega)},
\label{dpm}
\end{equation}
where  $\chi_{\pm}(q,\omega) = \Pi(q,\omega)/\epsilon_{\pm}(q,\omega)$.
The quasiparticle energies of the screened longitudinal vibrations are the
poles of $D(\omega,q)$. The poles of the $D(\omega,q)$ are complex and 
hence we need an analytic continuation of the response function. 
However, it is a good approximation

\begin{figure}
\epsfysize=2.in
\centerline{\epsffile{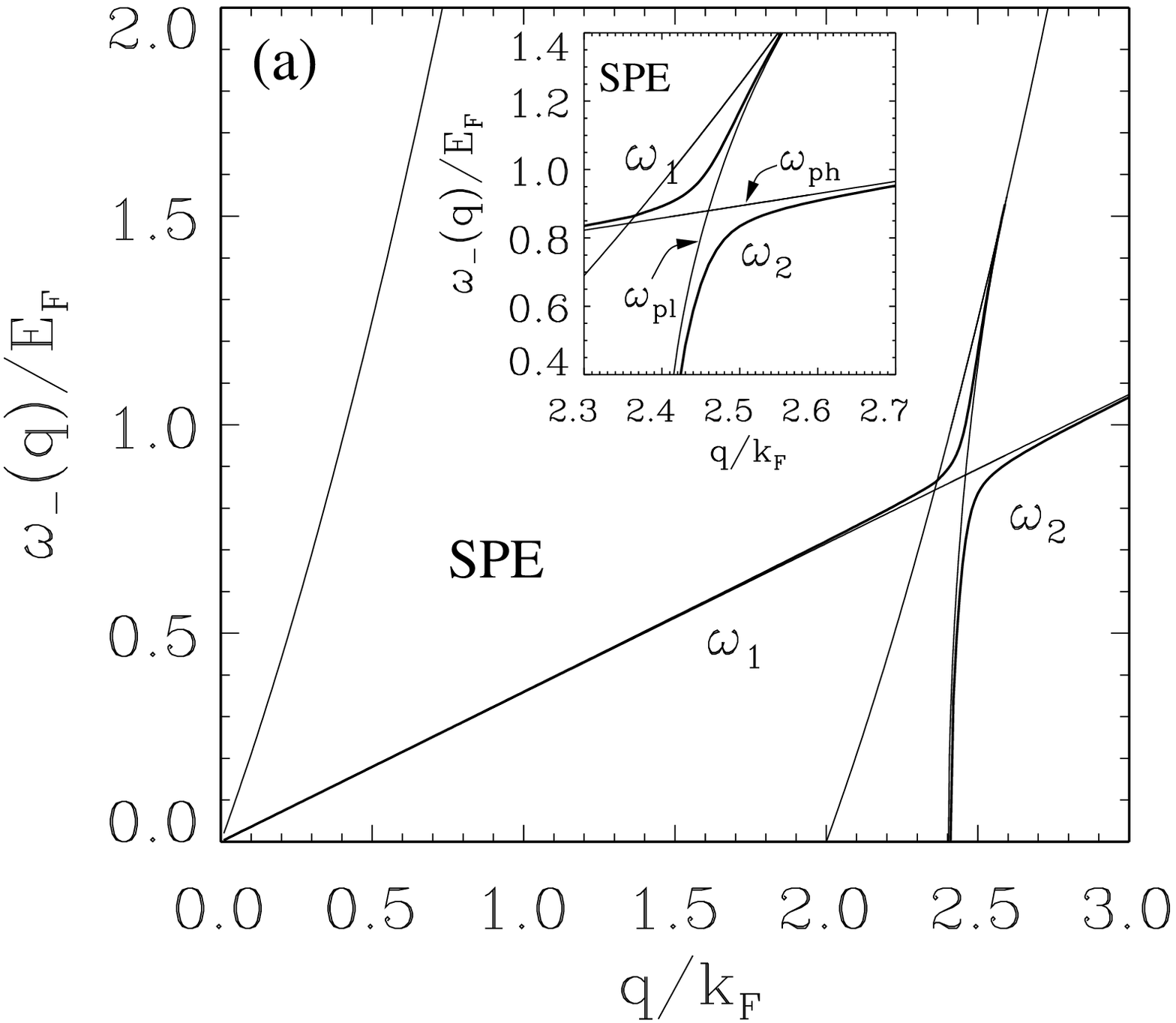} }
\vspace{0.5cm}
\epsfysize=2.in
\centerline{\epsffile{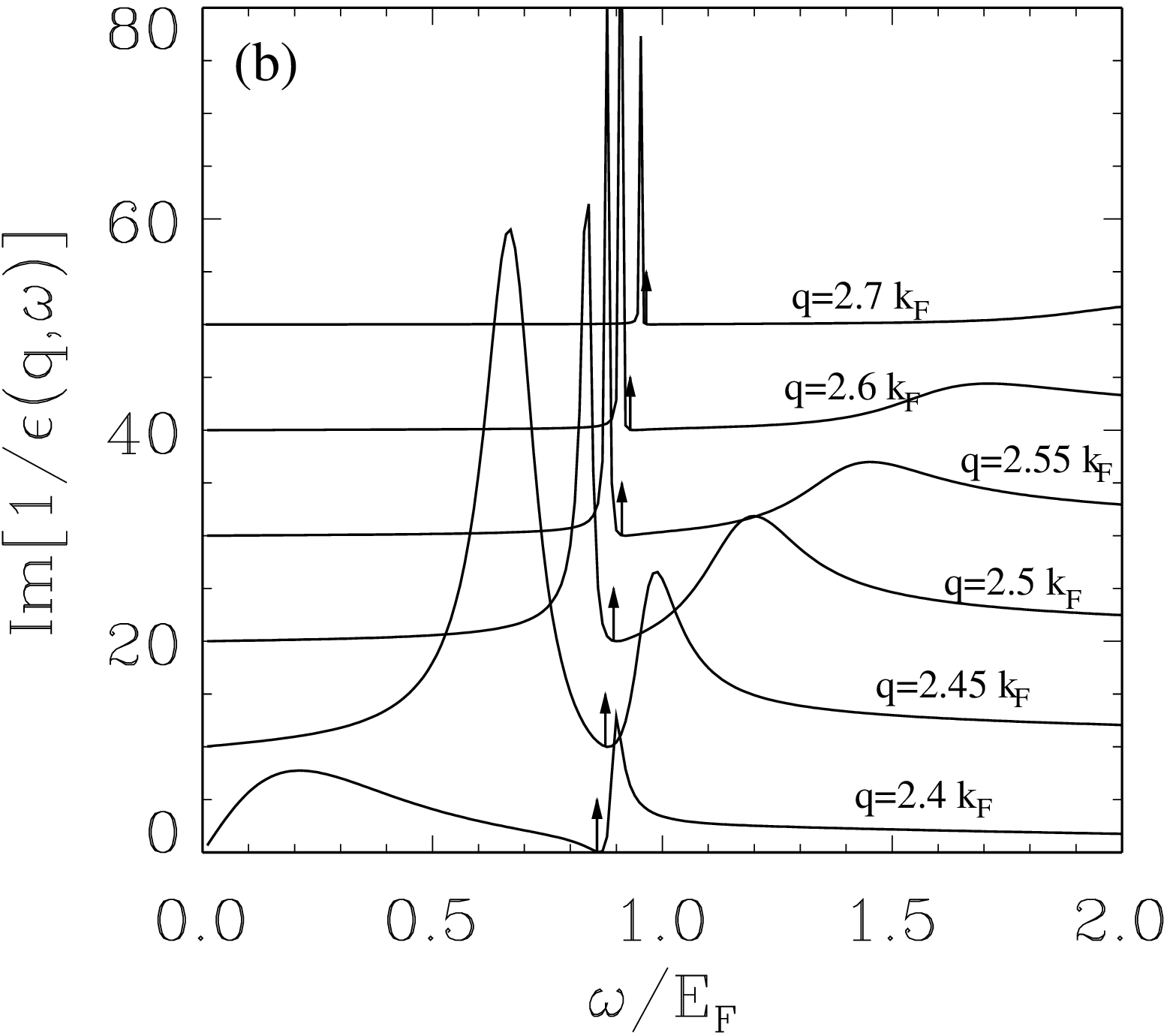}}
\vspace{0.5cm}
\caption{
(a) The coupled plasmon-acoustic phonon modes and
(b) the loss function (or the spectral weight) of the coupled mode, 
Im$[1/\epsilon^{tot}_-(q,\omega)]$, 
as a function of energy for fixed wave vectors. 
Inset in (a) shows the coupled mode near the AP-acoustic phonon resonance.
The area bounded by the light solid lines indicates 
the SPE Landau damping continuum.
Arrows in (b) are the energies of the bare phonon mode at given wave vectors.} 
\label{fig1}
\end{figure}

\noindent
to take the static limit 
($\omega = 0$) in the dielectric function and to 
replace $\omega$ 
by the real frequency $\Omega(q)$ in the imaginary part of the 
response function. Then we have
the renormalized phonon modes $\Omega_{\pm}$ and damping of the 
phonon $\gamma_{\pm}$:
$\Omega_{\pm}^2(q)  = \omega_q^2 - 2\omega_q M_{\pm}^2 {\rm Re}
\chi_{\pm}(q,\omega=0)$, 
$\gamma_{\pm}(q,\omega)  =  M_{\pm}^2 {\rm Im}\chi_{\pm}
(q,\Omega_{\pm}(q))/
[\omega_q-2M_{\pm}^2 {\rm Re}\chi_{\pm}(q,\omega=0)]$.
The width of the resonance in $D_{\pm}(q,\omega)$ is given by
$2\gamma_{\pm}(q,\omega)$.

In Figs. 1 and 2 we show the results for
the out-of-phase charge density modulation of the two layers in the 
bilayer system (the in-phase modulation, which is not shown, is
similar to the single layer case).
We use parameters  corresponding to electrons in 
GaAs quantum wells \cite{senna}: $m= 0.067 m_e$, 
$\kappa = 12.5$, $c_0 = 5.14\times 10^5$ cm/s, $\rho = 5.3$ g/cm$^3$, and
$D=-16.0$ eV for all our results. 
In Fig. 1 we show the coupled plasmon-acoustic phonon modes and
the loss function, Im$[1/\epsilon^{tot}(q,\omega)]$ which gives the 
mode spectral weight, as a function of energy 
for fixed wave vectors near the mode coupling region.
In Fig. 1 we choose $r_s=10$ corresponding 
to the electron density $n=0.33 \times 10^{10}cm^{-2}$,
the well width $a=100 \AA$, and layer separation $d = 300 \AA$. 
The area bounded by the light solid lines indicates the
SPE Landau damping continuum. 
When correlation effects are turned off (i.e., $G\equiv 0$)
we have a well defined acoustic 
plasmon mode above the electron hole continuum and 
a well defined acoustic phonon mode with the bare phonon 
dispersion $\omega_{ph} = c_0 q$. 
There is essentially no mode coupling
without correlations because of the large
difference in the plasmon and phonon energy scales. 
Local field effects give rise to a 
suppression of the AP mode above the SPE and 
its reappearance at large wave vectors
beyond the single particle continuum in spite of it being 
overdamped (by Landau damping) at long wavelengths. 
This reappearance
of the AP mode at large wave vectors, which is purely a local
field effect, produces strong plasmon-phonon mode coupling. 
In Fig. 1(b) we show the spectral weight of the coupled modes. We find 
two peaks corresponding to the coupled
modes, one below the bare acoustic phonon mode and the other above.
Since the in-phase plasmon mode $\omega_+(q)$ is essentially unaffected by
correlation effects and has much higher energy than the phonon mode,
the coupling of the in-phase plasmon mode to the acoustic phonon mode
is negligible. (This is also the reason why the phonon softening 
phenomenon predicted in this Letter cannot occur in single layer 2DES.)

In Fig. 2 we show the renormalized acoustic phonon mode dispersion and 
the damping for different layer separations 
by finding the complex poles of the 
renormalized phonon propagator, Eq. (\ref{dpm}). 
The renormalized phonon mode is 
very different from the bare phonon mode due to 
correlation induced mode coupling.
At the finite critical wave vector 
$q_c=2.41k_F$ for $d=300 \AA$ and $q_c=1.75k_F$ for $d=690 \AA$ we find 
phonon softening, where the acoustic phonon mode is renormalized to
zero frequency $\omega(q_c) =0$.
The critical wave vector depends on both density and layer separation,
since it is a correlation effect. We emphasize 
that the bilayer phonon softening takes place 
due to the resonant coupling of GaAs acoustic phonons to
the out-of-phase charge density modulation associated with the
bilayer acoustic plasmons, which can become very strong
at low enough carrier density. Strong acoustic 
phonon - acoustic plasmon coupling is the mechanism causing
the bilayer phonon softening. Our predicted phonon 
softening thus necessarily requires both low carrier density 
(large $r_s$) and a bilayer system.

We conclude by briefly discussing possible experimental means to observe
the predicted phonon softening. The most direct method would be to 
measure the bilayer phonon dispersion using ballistic phonon 
spectroscopy \cite{hensel} as was done for single layer 2D systems 
some years ago. The other possibility is to look for phonon 
softening indirectly through bilayer drag measurements \cite{gram} 
where the magnitude of the drag should develop interesting anomalies 
in the soft mode regime since the drag is mediated partly by the phonon 
exchange mechanism. It may also be possible to see the softening in 
bilayer transresistivity measurements \cite{kata}. In principle, it 
may be easier to see the softening phenomenon in hole bilayers where 
very large 

\begin{figure}
\epsfysize=2.5in
\centerline{\epsffile{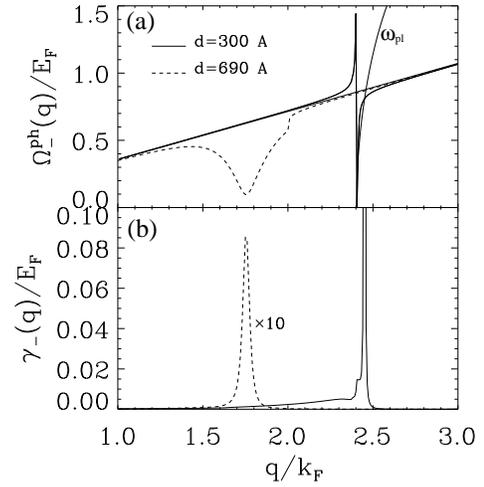}}   
\vspace{0.5cm}
\caption{(a) The renormalized phonon modes $\Omega_{ph}(q)$ and (b) 
the damping of the phonon mode $\gamma_-(q,\Omega)$ as a function 
of wave vector for  layer 
separations $d=300 \AA$ (solid lines) and $d=690 \AA$ (dashed lines)
with $r_s = 10$.
We show the plasmon (AP) mode $\omega_{pl}$ corresponding to $d=300 \AA$
for comparison.
}
\label{fig2}
\end{figure}

\noindent
$r_s$ values ($r_s \approx 20-30$) can be achieved.

This work is supported by the U.S.-ARO and the U.S.-ONR.


\begin{thebibliography}{99}

\vspace{-1.5cm}

\bibitem{ei} See, for example, the articles by J. P. Eisenstein and 
by S. M. Girvin and A. H. MacDonald in {\it Perspectives in Quantum 
Hall Effects}, edited by S. Das Sarma and A. Pinczuk 
(Wiley, New York, 1997).

\bibitem{das} S. Das Sarma {\it et al}., \prl {\bf 79}, 917 (1997);
A. Pellegrini {\it et al}., {\it ibid.} {\bf 78}, 310 (1997).

\bibitem{spi} I. B. Spielman {\it et al}., \prl {\bf 84}, 5808 (2000);
A. Stern {\it et al}., {\it ibid.} {\bf 84}, 139 (2000).

\bibitem{neilson}D. Neilson {\it et al.}, Phys. Rev. Lett. {\bf 71},
4035 (1993); \prb {\bf 44}, 6291 (1991).

\bibitem{hu}K. Flensberg and B. Y. K. Hu, Phys. Rev. Lett {\bf 73},
3572 (1994); \prb {\bf 52}, 14796 (1995).


\bibitem{dsh} S. Das Sarma and E. H. Hwang, \prl {\bf 81}, 4216
(1998); G. Gumbs and G. R. Aizin, Phys. Rev. B {\bf 51}, 7074 (1995).

\bibitem{dassarma}S. Das Sarma and A. Madhukar, Phys. Rev. B {\bf 23},
805 (1981). 

\bibitem{pinczuk} A. Pinczuk {\it et al.}, Phys. Rev. Lett. {\bf 56},
2092 (1986); G. Fasol {\it et al.}, {\it ibid}. {\bf 56}, 2517 (1986);
D. S. Kainth {\it et al.},
Phys. Rev. B {\bf 57}, R2065 (1998).

\bibitem{gruner} G. Gr\"{u}ner, Rev. Mod. Phys. {\bf 60}, 1129 (1988).

\bibitem{senna} J. R. Senna and S. Das Sarma, \prb {\bf 48}, 4552 (1993).

\bibitem{tanatar} B. Tanatar and D. M. Ceperley, Phys. Rev. {\bf 39},
5005 (1989).



\bibitem{hensel} J. C. Hensel, R.C. Dynes, and D. C. Tsui, \prb {bf 28},
1124 (1983).

\bibitem{gram} T.J. Gramila {\it et al.}, \prl {\bf 66}, 1216 (1991);
K. G\"{u}ven and B. Tanatar, \prb {\bf 56}, 7535 (1997).

\bibitem{kata} Y. Katayama {\it et al.}, \prb {\bf 52}, 14817 (1995).


\end{thebibliography}
\end{document}